\documentstyle[preprint,tighten,aps,psfig]{revtex} 

\newtheorem{definition}{Definition}

\newtheorem{proposition}{Proposition}

\newtheorem{properties}{\sc Property}

\newtheorem{corollary}{Corollary}

\newtheorem{exemple}{\sc Exemple}

\def\be{\begin{equation}}

\def\ee{\end{equation}}

\def\bea{\begin{eqnarray}}

\def\eea{\end{eqnarray}}

\def\bdf{\begin{definition}}

\def\edf{\end{definition}}

\def\bpr{\begin{properties}}

\def\epr{\end{properties}}

\def\bpt{\begin{proposition}}

\def\ept{\end{proposition}}

\def\bcll{\begin{corollary}}

\def\ecll{\end{corollary}}

\def\bex{\begin{exemple}}

\def\eex{\end{exemple}}

\def\d{{\rm d}}

\begin{document}

\preprint{\vbox{\baselineskip=12pt
\rightline{ICN-UNAM-00/07}
\rightline{gr-qc/0010086}
}}

\draft

\title{When is S=A/4? }

\author{
Alejandro\ Corichi${}^{1,2}$\thanks{E-mail:
corichi@nuclecu.unam.mx} and
 Daniel Sudarsky${}^1$\thanks{E-mail: sudarsky@nuclecu.unam.mx} }
\address{1. Instituto de Ciencias Nucleares\\
Universidad Nacional Aut\'onoma de M\'exico\\
A. Postal 70-543, M\'exico D.F. 04510, M\'exico.}
\address{2. Department of Physics and Astronomy\\
University of Mississippi, University, MS 38677, USA.}

\maketitle

\begin{abstract}

Black hole entropy and its relation to
the horizon area are considered. More precisely,
the conditions and
specifications that are expected to be required for the assignment
of entropy, and  the consequences
that these expectations have when applied to a
black hole are explored. In particular, the following questions
are addressed:
When do we expect to assign an entropy?; when are entropy and area
proportional? and, what is the nature of the horizon?
It is concluded that our present
understanding of black hole entropy
is somewhat incomplete, and  some of the relevant
issues that should be addressed in pursuing these questions are
pointed out.


\end{abstract}


\section{Introduction}

In the past 25 years there has been a great  deal of activity
around the nature and origin of  black hole entropy, since the
pioneer work of Bekenstein and Hawking who found a close relation
between entropy and area of black holes \cite{1,2,haw:cmp}. It
is generally regarded that the identification of the fundamental
degrees of freedom and the computation of the entropy for
a black hole is one of the
major challenges for any candidate quantum theory of gravity. Of
particular importance are the recent attempts to recover, from
basic approaches to quantum gravity, the ``standard expression" for
the entropy of a black hole \cite{string,abck}. For a recent
review see \cite{wald:review}. Our aim in this article is to
review and discuss the foundations of those attempts by examining
in detail under what conditions this standard answer can be
expected to be obtained.

The question of when entropy is equal to $A/4$ for a black hole is
in fact  made of at least 3 questions:

\begin{enumerate}
\item\label{1} To what exactly and under what conditions
 do we expect, in general,  to  assign an entropy?

\item\label{2} When is the entropy assigned to a
black hole equal to 1/4 of its area?

\item\label{3} To which area exactly do we refer in
answering the above question?

\end{enumerate}

The first (and probably the most controversial) of
these questions is in itself a collection of questions:

\smallskip

1.a) Under what physical circumstances do we assign an entropy?
\smallskip

1.b) To what do we assign entropy?: To a history?, to an ``instantaneous"
 state of a system?, to a our description of such a state of the system?
\smallskip

1.c) What needs to be specified in order to assign an
entropy?

\medskip

More concretely, question 1.a), namely under what circumstances
 we assign entropy,
 refers to the issue of assigning
entropy to: i) all situations, ii) stationary situations, iii)
quasi-stationary situations, or iv) some other set of situations,
larger than ii) but smaller than i).

In answering this question, we are also led to the issue of what
type of entropy are we talking about: thermodynamic entropy or
statistical mechanical entropy. Within this later case we should
also decide whether we refer to the Boltzmann or the Gibbs
entropy.

Once one has decided when to assign entropy, one should also specify
to what ``object" it should be associated. For instance, one
should say if the entropy is to be associated with the interior, the
exterior or the horizon of a black hole.
Finally, the question of what needs to be specified in order to
assign entropy
[(1.c)], refers to standard specifications that one provides
in treating the statistical mechanics of any system, for instance,
the distinction between system and observer, coarse graining, etc.

On a closer look, one can see that all the questions that
make up the question of when to
assign entropy, are in a sense, an indication that we need to put
the concept of entropy on a firmer ground than it is at the moment.
 It might seem that the fact that  statistical mechanics can deal
successfully with all practical applications of the subject is an
indication that all is well and
clear. However,
to understand the limitations let us focus for a moment on another
concept where the situation is fully clarified, but was not so when the
concept was firstly used:
The concept of energy.  For a start, one realizes that for free particles
the kinetic energy is conserved, and then, that for certain forces one
can introduce a potential energy, where now it is the sum of the  former
and the latter that constitutes the true conserved energy. The story can
be continued  with the realization that one needs to incorporate more
forms of energy in order to  maintain the validity of  the conservation law
in more general circumstances.  It is only when one regards energy as the
``conserved quantity in the presence of time-invariance", that the
mystery behind all its apparent manifestations disappears, and one
recognizes that one is truly talking  in all cases about one
single quantity: energy.
Similarly, it seems that we would need an `unified' definition of
entropy similar to the one available for energy, in order
to clarify the  situation and in particular  the
discussion around question 1, namely the question of when
to assign entropy to a system. In other words, in the case of energy one
knows that one of the implied aspects of having a single energy
conservation law, instead of, say, one for kinetic energy and one for
potential energy, is the tacit acknowledgment that there exits physical
process  that  convert one type of the energy into another.
Moreover, given two types of energies, one can
in principle (in  almost all  cases) find one physical
process taking the first type into the second. The exceptions are in fact
supposed to be codified trough the concept of   entropy and by the
second law of thermodynamics (besides the usual limitations arising from
 other conservation laws). In the same way, if such a `unified recipe'
to deal with entropy was at hand, one would hope to understand not only
the second law (in its generalized form including, in particular,
 the entropy of  black holes), but also the limitations, if any,
to converting one type of entropy into another.

 This last point is
particularly interesting, because as far as the authors
 know,  no such restriction
has been advanced to date and thus the second law is the only
limitation for the conversion of energy from one type to another.
In fact one can give strong arguments that such restrictions must
exist, in particular, associated with the issue of locality: No
one would believe that one is allowed to device a machine that,
say, transfers heat from a cold reservoir to a hot one without any
additional local effects,  and which avoids violating the second
law simply by having  a second part of the machine that creates
sufficient entropy in a distant galaxy. Such considerations
clearly indicate that one must face the issue of localization of
entropy in general, and by having a black hole as a part of such
contraption, the localization of black hole entropy in particular.
One would like to have a general definition, analogous  say to the
definition of the energy momentum tensor,  of an entropy current
$S^a$ (see \cite{wald:review}) satisfying an equation of the form:
\be \nabla_aS^a \geq 0 \ee such that the entropy associated with
an hypersurface $\Sigma$, with unit normal $n^a$, and volume
element $\d V$,  given by,
\be
 \int_{\Sigma}  n_a S^a \, \d V\, ,
\ee
be
such that the entropy associated to an hypersurface is greater or
equal to that associated with an hypersurface to the past of the
first one.

 Such a general and precise notion of entropy would seem to be
required, before one can hope to have a complete understanding of the
reasons behind the validity of a  generalized second law (including the
contribution  to entropy associated with black holes). There are in fact
proposals for the derivation of this generalized second law\cite{Sorkin1},
 which however fail to clarify the
underling reason for their validity, in the sense that, for
example, the conservation of energy
is understood as a result of a time invariance of the
underlying theory. Needless to say, these issues can not
be treated with the current understanding, because we lack, among other
things, the notion of localization of the entropy.

Assuming that one is willing to consider entropy as assigned to
all situations, one is then confronted with  question 2: is the
entropy associated to a black hole always proportional to its
area? It is clear that in (quasi)stationary situations, the
existence of the first law leads us to the conclusion that entropy
{\it is} proportional to area. Then, the question narrows to:
should we consider area as a measure of entropy, also for the
dynamical case?

The third question then refers to the identity of the area
horizon one
wishes to consider in order to identify it with entropy.
There are situations in which all known notions of horizon agree
(if defined).
In particular, this is true when the spacetime under consideration
is stationary.

However, already in
the quasi-stationary case there are differences between, say, the
event horizon and an isolated horizon. In the dynamical case, none
of the definitions agree. It is then of vital importance to
``make up our mind" about the nature of the relevant horizon.

It is important to stress that the question of when entropy and area
are proportional (question 2) and to the nature of the horizon
(question 3) have to be considered
once one has tried to answer question 1 in detail. In the
hypothetical case that one has an ``universal" definition for
entropy, one might hope that question 3 would be settled `ab initio'
and that question 2 can be answered by a direct application of
the relevant formalism (assuming one has a quantum theory of gravity).

The purpose of this work is to critically review our current
understanding of the foundations of black hole entropy
and to point out the interrelations (not always
fully appreciated) between the positions one takes in
answering the various questions here posed,
and the requirements that   the positions one adopts in facing
each of these issues ought to be mutually consistent.
 It is
important to stress that this article does not intend to give a
global answer to the issues that are addressed, but rather to
point out the unresolved issues.

In this work we restrict  our attention to the
general theory of relativity (with, in principle, arbitrary matter
couplings), and do not consider higher derivative theories, for which
the relation between entropy and area does not seem to hold
even at the classical level
(i.e., in the generalized first law)\cite{Vivek}.

This paper is organized as follows: In Sec.~\ref{sec2} we
discuss the question of when entropy is defined. Section~\ref{sec3}
is devoted to the study of when entropy and horizon area are
proportional. The question of the nature of the horizon is the
subject of Sec.~\ref{sec4}. Finally, we end with a discussion
in Section~\ref{sec5}.

\section{When is entropy defined?}
\label{sec2}

First, let us elaborate on the question of what kind of
entropy we should focus on,
namely thermodynamical vs statistical.

We know that the thermodynamic entropy is associated with
stationary and quasi-stationary situations.
In the case of black holes, this is normally reflected in the
existence of the ordinary first law, and the thermodynamic
entropy would be the quantity appearing there.
Nevertheless, we are in fact interested in the statistical mechanical
entropy because, as one can argue, it is the most general kind of entropy
since, for every situation in which
the thermodynamical entropy is  defined, so is the statistical mechanical
entropy. In these cases
the two essentially agree, but there are situations is which the
(standard) thermodynamic entropy is not even defined.
Furthermore, it is the
statistical mechanical entropy the object that, in principle,
can be calculated from the basic microscopic theory, which
in  the case of a black hole's entropy  contribution   would be the
quantum theory of
gravity\footnote{It is also known that from purely dimensional reasons,
one would need to
introduce Planck's constant $\hbar$ in order to identify entropy with area
\cite{1}. This
expectation was fully realized in Hawking semi-classical calculation
 \cite{haw:cmp}, thus
indicating that black hole entropy is quantum mechanical in nature.}.

Now, the two types of statistical entropy, namely Boltzmann and
Gibbs are, in principle, conceptually different. The first,
depends on the exact microstate of the system under consideration
and is defined as the logarithm of the number of microstates
being ``macroscopically indistinguishable'' from the given one.
This set is said to represent the mesostate. That is, if $N_i$
denotes the number of microstates making up the $i^{th}$
mesostate, then the Boltzmann entropy is,
\be
   S= \log N_i
\label{Boltz}
\ee
whenever the microstate finds itself within the $i^{th}$ mesostate.

The second is an ``ensemble functional'', rather than
a function of the actual physical state of the
system.  Associated with the ensemble there is a probability density
$\rho$ on the space of {\it microstates}, and the Gibbs entropy is
\be
    \int \d x\, \rho(x) \log (1/\rho(x)) .
\label{Gibbs}
\ee

However, in practice  we use an essentially identical coarse
graining prescription to define the level of uncertainty that
allows us to construct either the notion of mesostate for
Boltzmann entropy or the ensemble for the Gibbs entropy. Since the
statistical mechanical entropy is more general than the
thermodynamical entropy, the former must be defined at least in
stationary and quasi-stationary situations [ii) and iii) above].
We know of no criteria that would
be appropriate to specify a more general situation (case iv) above)
 so our options for when the statistical
entropy is defined seems to be restricted to just the set i) (i.e.
always) or the same as the thermodynamical entropy (i.e.
stationary and quasi-stationary cases).

We would like to further argue against the choice of
having the statistical entropy defined only in the stationary
and quasi-stationary cases. Making this choice
would render the second law as
practically useless to prevent, for instance,
the construction of perpetual motion machines ( i.e.
 consider a contraption that moves in
such a way as to avoid passing trough a situation for which entropy is
defined).

Moreover, this option would  seem to destroy the Markovian nature
of a physical theory, namely the property that the predictive
power of a physical theory is not increased by considering the
past together with the present, as compared to the consideration
of  the present alone.   That is, if the present corresponds to a
situation in which entropy is not defined, we could not preclude a
future situation with a given entropy $S_1$, in terms of our
knowledge of the present, but could, for example,  preclude that
situation if we use (together with the second law) the fact that
in the past the (closed) system was in a situation characterized
by an entropy $S_0>S_1$. One could argue that the Markovian aspect
should be associated with the whole physical theory and not with
each physical law separately. However,  in this case,  the fact
that the second law is the only law incorporating a  particular
arrow of time,  it seems difficult to imagine that the other known
physical laws could restore  the Markovian nature to the situation
at hand. All these considerations seem to take us to the position
of accepting 1.a.i), i.e {\it The statistical entropy should be
defined for all situations} (See also \cite{StatMechDyn}).

Now, in providing this answer we are in fact providing also
a partial answer to the question of which object should one assign
an entropy to
[1.b) above] in the sense that we are assigning an entropy to a
physical situation with a notion of localization in time.
The alternative of assigning an entropy to a complete spacetime or a
history (as indicated for example in \cite{hawking2})
would seem to make the concept of entropy completely useless
in the sense of adding predictive power, and in particular as a tool
for ruling out perpetual motion machines. In a sense we are puzzled as
to what would be the current
view of the authors in \cite{hawking2} about the area
increase theorem in classical general relativity and its
resulting connection with the generalized second law.

Another, more moderate view, would be to assign an entropy with  a
situation
localized in time but within the context of a spacetime or history.
Here again  there are potential problems if we require the history to be
known to a larger extent that is possible from a description in
terms of initial
data -a situation  that could occur if we let quantum events play a
decisive role in the selection of
the possible histories, as in the examples discussed in \cite{we}-
as we could render the concept
of entropy, again,  useless in ruling out perpetual motion machines.
These problems would be avoided
if we accept that there be
no supplementary information in the history.

Thus, from the previous discussion
we are lead to the conclusion that  in order for the concept
of entropy to be useful
in the ways we expect it to be,  {\it we must assign entropy under
all circumstances and to instantaneous physical situations}.

The remaining aspect of question 1.b), namely to
which object to assign entropy is in fact incorporated within the
question of what needs to be specified in order to define entropy
[question 1.c) above] which we address  next.
It is obvious that we must specify at least the
physical system to which one is about to associate an entropy, and
 the coarse graining needed to define the
macrostate (in the Boltzmann scheme) or the ensemble (in the Gibbs
scheme). This leads to the conclusion that in fact we do not
assign entropy to a physical situation, but {\it to
our description of the physical situation}.
This view is consistent with the assignment of entropy to, say, de Sitter
and Rindler horizons as in \cite{DeSitterRindler}.
This is a  consequence of the fact that
there is, in principle, no natural specification of the coarse graining,
if no specification is given of the
experimental procedure  used to prepare the system.
Thus, entropy would seem to be
a relative concept, with different
physicists disagreeing about the value of the entropy of a specific system
at a given ``time". This in itself  is  not so worrisome,
after all,  other concepts like energy or
length suffer from the same relativism and are nevertheless very useful.
However, this indicates that we must specify the
observer with respect to which entropy is to be assigned.
It is even possible that the two specifications, observer and
coarse graining become intertwined, as for example
in the case of a Black Hole, for which one
can think that the prescription to disregard whatever occurs inside
is in fact part of the coarse graining \cite{we}, or part of
the specification of the observer
(whom we can think is restricted to move in the outside thus
having no access to any information about  the inside).
This must not be interpreted as in any way saying that the entropy of
a black hole, is associated
with the internal degrees of freedom
 since that would lead to  various problems \cite{Sorkin1,jacob},
but only as the statement that
there is entropy to be assigned in this case because we are disregarding
the inside. That is, we leave
open the possibility that, for instance, the entropy might have to do with
those degrees of freedom of the inside which can
affect the outside through boundary conditions, or correlations, which
 would lead to the
view that the relevant degrees of freedom are those associated with the
horizon \cite{jacob}. Returning to the issue raised in 1.c), in
particular, to the details of the
specification of observer and its observational capacity, we must,
in accordance with the limitations imposed by the relativistic
nature of physical reality, agree from the onset that the observer
must be replaced by a collection of observers.
We are
faced then with the problem of having to specify the extent of
this set of observers in order to be able to decide the extent of
the observable quantities, a specification that one could hope
will take into account, for example, the existence of horizons of
various sorts. In particular, in the case one  wished to consider
event horizons in the previous discussion one would immediately
run  again into the  problems derived from the teleological nature
of such object, i.e. the fact that the present location of the
horizon depends strongly on  events in the future that might, as
in  the example considered in \cite{we}, make it  impossible to predict its
location
based on the  full knowledge of the system at present.

\section{Stationary vs. Dynamic}

\label{sec3}

Now we turn to question 2. That is, when is the entropy of a black hole
proportional to its area?
Here again we seem to face various alternatives:
2.a) In  stationary and quasi-stationary situations; 2.b) Always; and 2.c)
Some other restricted set of circumstances.

If we take the option that entropy and area are proportional only in
stationary and quasi-stationary situations,
we immediately face two questions. First,
what are we going to take as the
expression for entropy in other situations?
One can try to answer this question both at the classical level and
in the quantum domain. Classically, one would have to define a
geometrical quantity to be associated with a dynamical entropy.
Several such attempts are available in the literature
\cite{Vivek,jkm,hayward}.
Note that the particular prescription becomes intertwined with
our question 3, that is, with the nature of the horizon one wishes to
consider.
A second possibility is that the answer will be available
only when we have a full theory of quantum gravity.
After all, if asked to compute the entropy of a
given non-equilibrium  (macroscopic) configuration of a mass of gas, we
need to go to the microscopic
theory to count microstates etc, and we can not expect the answer to be a
simple function of a single
macroscopic parameter (which might not even be defined).
Furthermore, within a full quantum theory of
gravity, a generic configuration might not even have a macroscopic
description
in terms of a space-time with a horizon in it (like in some of the
D-brane calculations \cite{string}).

An important issue in both cases is, in a sense, the other side of the
same coin, i.e., what are
we going to make of the area theorem? If the dynamical entropy is not
to be identified with the area of the event horizon, the existence of
this theorem will lead to a strange situation because we will have on the
one hand, the second law for
the true entropy (which would not be a simple expression) and the
second unrelated  non-decreasing
quantity: the area of the event horizon.

Furthermore, we note that the first law of black hole mechanics:
\be \delta M ={\kappa \over {8 \pi}}\, \delta A + \delta W \ee
where $M$ is the ADM mass, $\kappa$ the surface gravity, $A$ the
area of the event horizon, and $\delta W$ stands for work terms,
is known to be valid for arbitrary  variations of stationary black
holes \cite{SW} even if these configurations are unstable. Thus,
the point of the phase space
to which  the variation  has taken the original stationary black
hole, is not only not stationary, but  it can not be said to be  a
configuration that would remain close to a stationary one, i.e.
can not be said to be quasi-stationary.  Nevertheless the
identification of this law with the first law of thermodynamics,
clearly   indicates that we are assigning an entropy   $S\sim A$
to these  black holes.

Assuming that we take option 2.b), namely that the entropy is
always proportional to the horizon area, then we are lead to a
problematic situation because of the fact that we need to know the
area of which horizon we are talking about. The event horizon
seems not to be adequate since we need to know the complete
spacetime in order to locate the horizon, and, as we concluded
previously, the entropy need to be assigned to an instant of time,
which in general relativistic settings corresponds to Cauchy
hypersurfaces. We could take the view that the prescription is,
then, to take the data associated with the hypersurface, including
the geometry and the matter fields, evolve it according to
Einstein's equations and proceed to locate the horizon in the
corresponding hypersurface. Unfortunately this is also not a
viable option in general as demonstrated by the example discussed
in \cite{we}, in which the initial data, although complete, is not
enough to locate the horizon on account of a decisive role played
by a quantum measurement that is to be performed to the future of
the given hypersurface, leading to fluctuations of the area of the
event horizon with \be \Delta A \approx A. \ee

The previous discussion  seems to lead us to option 2.c), that is,
to some other restricted set of circumstances, which still needs
to be specified. In this regard, again the analysis of the example
in \cite{we} would point to the following generalization of the
previous prescription: take the data associated with the
hypersurface, including the
 geometry and the matter fields,  consider the possible evolutions
taking into account quantum alternatives and proceed to locate the
horizon on the initial hypersurface for each of the corresponding
spacetimes, and finally add the corresponding values of the areas
with the appropriate probabilistic weights. So far we have
centered our discussion of this question assuming that the entropy
would be associated with the area of the event horizon, and in
fact, the alternative 2.c) (some other restricted set of circumstances)
 is also an opening for the consideration of the next section.

\section{Area of What?}
\label{sec4}

In the past sections we have argued that the event horizon, even
when it has a clear space-time definition, and is in a sense the
obvious choice one might make, has several problems for a
satisfactory definition of entropy. Again, the main problem of
choosing the event  horizon is its teleological nature that makes
the situation {\it different} (as explained in Ref.\cite{we}) from
the case of an ordinary thermodynamical system  put in a quantum
superposition of states. As explained in \cite{we}, if one wishes
to adopt the event horizon, then one needs to give up a canonical
theory and/or modify the existing quantum theory. On the other
hand, if one is not willing to give up a canonical quantum theory,
then one can not consistently insist on  the event horizon as the
relevant quantity. One might conclude that, in this case, the
event horizon should be replaced by another geometrical quantity
in dynamical situations
 and then one is faced with the problem of finding a suitable
alternative. The purpose  of this section is to review the
available possibilities, which to our knowledge are: 1) The
apparent horizon \cite{penrose}, 2) The isolated horizon
\cite{abf,ack}, and 3) The trapping horizon \cite{hayward}.

The apparent horizon
 has very serious problems, since it is known to be discontinuous
for dynamical situations like a collapsing star
\cite{hawking:lh},\cite{we}. Furthermore, it is known  that
even the Schwarzschild spacetime contains Cauchy hypersurfaces
with no apparent horizons.

The second alternative, namely
isolated horizons, are particularly interesting for several reasons.
First, it has been shown that for {\it quasi-stationary} processes,
the (quasi-local) horizon mass satisfies a first law in which the
entropy is proportional to the horizon area \cite{abf}.
Secondly, there exists a calculation
of the statistical mechanical entropy 
that recovers the ``standard result" $ S= A/4 $ for various types
of black holes \cite{abck}. This formalism is in fact a
generalization of the standard stationary scenario to
more physically
realistic situations, because the exterior region need not to be
in equilibrium. Nevertheless, the whole approach is based in the
assumption that the horizon itself is in internal  equilibrium. In
particular, its area has to be constant, and nothing can ``fall
into the horizon''. In this regard, isolated horizons as presently
understood, are not fully satisfactory since the formalism is not
defined and does not work in general, dynamical, situations.

Moreover, there are situations in which one is faced with the
occurrence of several isolated horizons, intersecting a single
hypersurface, one within the other, and one must decide to single
out the one to which entropy is to be assigned. We can take the
view that this  should be the outermost horizon, but this seems to
be just an {\it {add hoc}} choice, unless it is argued that the
selection is the natural one associated with the fact that we are
specifying the ``exterior" observers  to be the ones with respect
to which entropy is assigned. This view would be
 natural if we take the position that the assignment of entropy  is
related to the coarse graining,
which is partially specified by pointing out the region from which
information is available to the
observer. However, this point of view would conflict with the fact that
the isolated horizons are
not good indicators of such regions, basically because their
definition is purely local and thus not
fully based on causal relations.

Isolated horizons are well defined for equilibrium situations. If
some matter or radiation falls into the horizon, the previously
isolated horizon $\Delta_0$ will cease to be isolated, and (one
intuitively expects) there will be in the future a new isolated
horizon $\Delta_1$, once the radiation has left and the system has
reached equilibrium again. One would like to have a definition of
horizon that interpolates between these two isolated horizons
$\Delta_0$ and $\Delta_1$, such that the physical situation can be
described as a generalized horizon that ``grows'' whenever matter
falls in. There is a natural direction for this notion of horizon,
and this leads us to the third possibility, namely, trapping
horizons \cite{hayward}.

In a series of papers,
Hayward has been able to show that there exists
(at least in the spherically symmetric case) a
dynamical (as opposed to quasi-stationary) first law, for a
(quasi-local) energy that, however,  does not coincide with
the horizon energy of the isolated horizons formalism (in the
static limit). There
exists also a second law, for the area of the trapping horizon, when
a particular foliation of the space-like horizon is chosen.
However, we face the problem that,  by definition,
these horizons can be specified only when the full spacetime is available:
given a point in
spacetime, the issue of whether or not it lies on a marginally trapped
2-surface, can not in general,
be fully ascertained until the whole spacetime (where the rest of the
2-surface is to be located) is
given. These option is also problematic because the trapping horizons are
in general space-like and
thus there is no guarantee that a given hypersurface would not
intersect the horizon in several components thus leading to the same
problem of in-definition
that was
mentioned in connection with option 2). Moreover, in this case the horizon
can even
be tangent to the hypersurface which  is an extreme version of the
previous problem.  The fact
that
all this  objections can be raised against this option, has its origin in
the fact that the
trapping horizon is not a surface defined on the grounds of causality alone.

It would be interesting to have a coherent
formalism developed, which incorporates
both the isolated and the trapping horizons formalisms, and that
allows for a dynamical description of ``black hole horizons''.

\section{ Discussion and Conclusions}

\label{sec5}

\bigskip

In this paper, we have critically reviewed our current
understanding of black hole entropy, focusing on the conceptual
foundations that lie below the assignment of entropy to a black
hole. In particular, we have  argued that for  the full content of
the second law to be useful one needs to take the view that
entropy should be assigned to all physical situations. Moreover we
also  argued that entropy should be assigned to situations
localized in time, which in the general covariant setting implies
that it should be assigned to
 Cauchy hypersurfaces, which can be
viewed  as immersed in  spacetimes but only to the extent that the
spacetimes themselves can be
obtained from the data on the corresponding hypersurfaces.

The conclusions above seem to be very robust in the sense that
taking an alternative viewpoint
would seem to force us to  deprive the second law from its predictive
power and therefore  from much of its meaning.

This has lead us to a rather paradoxical situation when dealing
with
 the entropy of a black hole,  because  the alternatives that are available
to  play the role of entropy in the general dynamical  case are
all suffering from serious disadvantages: The event horizon can
not in general be localized on a given hypersurface, and the best
that can  be expected is to have a probability for its various
possible locations  associated with the  various possible
spacetime developments of the given initial data. The apparent
horizon would lead to the assignment of zero entropy even to some
Cauchy hypersurfaces of the Schwarzschild  spacetime. The Trapping
horizons are also in general not localisable in the absence of the
full spacetime and moreover are in general space-like, a feature
that can lead to the multiple  crossings of the horizon with the
given space-like hypersurface, unless a preferred foliation of the
horizon is chosen to begin with, and therefore a natural
definition of ``time evolution" along the horizon.

One possible conclusion is that the general  expression for the entropy of
a dynamical back hole that would arise from a complete quantum gravity
theory, would be  rather complicated and
dependent on the details of the  theory, a
situation which would put such entropy in a similar footing with  the
entropies of other systems
which, in the dynamical situation, are not expressible as simple functions
of a few macroscopic
parameters. The puzzling aspect of this view is the meaning we would
ascribe to the classical area
increase theorem, since we would have to abandon its interpretation as
being just an expression of the second law,  in the case of classical
black holes.

Another possibility is that, in the semi-classical limit of the
full theory, the  states  that
yield space-times with a horizon in it,
would have the property that the horizons are quasi-stationary (or
even isolated).
`Dynamical states' would be present but they might not be interpreted as
a classical space-time with a dynamical horizon. This drastic conclusion
would of course have deep implications on black hole evaporation
and information loss. This  would imply that there are no quantum
counterparts to classical dynamical black holes in certain regimes.

An alternative
(an more moderate) conclusion would be to take the
entropy as the probability-weighted average of the
 event horizon areas associated with the possible future developments of
the appropriate Cauchy data.
This was, for instance,  the viewed taken in the
analysis of \cite{we} where an argument in  favor of this
proposal was obtained from a sum-over-histories formulation  of quantum
mechanics,  together with
the hypothesis that taking $S =-{\rm tr}(\rho \ln \rho)$ with $\rho$ the
density matrix for the
exterior black hole region yielded the correct result
$A/4$ in the standard situations.

This discussion leads us to conclude that quite aside  the issue
of finding a correct theory of quantum  gravity,
the status and interpretation of the thermodynamics of
black holes  is
rather  incomplete. Furthermore, the basic issues raised in
its connection could prove fundamental as
guidance in the search for the quantum theory of gravity\footnote{Thus, in
contrast with the  stationary black hole situation,
for which there seems to be no clues distinguishing the
various approaches towards a theory of quantum gravity
(as has been argued for instance in \cite{Carlip} (however see
\cite{dreyer})), we
expect the dynamical black hole case to be a much more demanding,
therefore a more selective test for such proposals.}.

\medskip
\section*{acknowledgments}

We would like to thank A. Ashtekar for helpful comments. This work
was in part supported by DGAPA-UNAM Grant No. IN121298 and by
CONACyT grants J32754-E and 32272-E. AC was also partially
supported by NSF grant No. PHY-0010061.

\end{document}